\documentclass[10pt,preprint2]{aastex}

\def\thvec{{ \mbox{\boldmath $\theta$} }}
\def\vissvec{{\bf V_{\rm iss}}}
\def\viss{ V_{\rm iss} }
\def\svec{{\bf s}}
\def\kappavec{{ \mbox{\boldmath $\kappa$} }}
\def\thetavec{{ \mbox{\boldmath $\theta$} }}

\def\simless{\mathbin{\lower 3pt\hbox
   {$\rlap{\raise 5pt\hbox{$\char'074$}}\mathchar"7218$}}} %< or of order
\def\simgreat{\mathbin{\lower 3pt\hbox
   {$\rlap{\raise 5pt\hbox{$\char'076$}}\mathchar"7218$}}} %> or of order
\def\be{\begin{eqnarray}}
\def\ee{\end{eqnarray}}

\begin{document}

\title{Modeling of Interstellar Scintillation Arcs from Pulsar B1133+16} 
\author{\sc Frank S. Trang \altaffilmark{1}}
\affil{Department of Electrical \& Computer Engineering \\
University of California San Diego La Jolla, CA 92093-0407 \\
and Avaak, Inc., 5405 Morehouse Dr., San Diego, CA 92121}

\author{\sc Barney J.\ Rickett \altaffilmark{2}}
\affil{Department of Electrical \& Computer Engineering \\
University of California San Diego
La Jolla, CA 92093-0407}

\altaffiltext{1}{e-mail:iknowfrank@yahoo.com}
\altaffiltext{2}{e-mail: bjrickett@ucsd.edu}

%\maketitle

%---------------------------

\begin{abstract}

The parabolic arc phenomenon visible in the Fourier analysis of the scintillation spectra of pulsars provides a new method of investigating the small scale structure in the ionized interstellar medium (ISM).  We report archival observations of the pulsar B1133+16 showing both forward and reverse parabolic arcs sampled over 14 months. These features can be understood as the mutual interference between an assembly of discrete features in the scattered brightness distribution. By model-fitting to the observed arcs at one epoch we obtain a ``snap-shot'' estimate of the scattered brightness, which we show to be
highly anisotropic (axial ratio $>10:1$), to be centered significantly
off axis and to have a small number of discrete maxima,
which are coarser the speckle expected from a Kolmogorov spectrum of interstellar plasma density.  The results suggest the effects
of highly localized discrete scattering regions
which subtend 0.1-1 mas, but can scatter (or refract) the radiation by
angles that are five or more times larger.
\end{abstract}


%================================================

\section{Introduction}

\par\noindent
The scattering of the radio waves as they propagate 
through the ISM has long been a clue to small scale 
inhomogeneities in the ionized interstellar plasma.  
The scattering is seen in the dynamic spectrum of 
pulsar signals as random modulations in frequency 
and time.  Islands of intensity in the dynamic 
spectrum, often referred to as ``scintles'',  are 
sometimes modulated by a crisscross pattern.  By 
Fourier analyzing the dynamic spectrum, Stinebring 
et al.\ (2001) discovered the remarkable phenomenon 
of parabolic arcs that underlies the criss-cross
substructure.  In this paper, we analyze parabolic 
arcs from archival observations of PSR~B1133+16 
originally reported by Gupta, Rickett, \& Lyne (1994).

The two-dimensional Fourier spectrum of the primary 
dynamic spectrum is often referred to as the secondary 
spectrum. However, as described by Cordes et al. (2006), 
it can also be regarded as the ``differential 
delay-Doppler'' distribution, whose coordinates are 
time delay ($f_{\nu}$) and Doppler frequency (or 
fringe rate $f_t$), being the variables conjugate 
to radio frequency $\nu$ and time $t$.  The theory 
of the arcs is described in that paper, which 
henceforth we refer to as CRSC, and is also discussed 
by Walker at al.\ (2004).

The basic arc phenomenon is remarkably simple, even 
though it lay undiscovered for over 30 years.  When 
scattered waves arrive at the observer from two angles 
they interfere and cause fringes in intensity.  The 
observer is moving relative to this interference 
pattern and sees a sinusoidal variation in both time 
and frequency, which thus appears as a pair of points 
in the secondary spectrum. The coordinates of these 
points are the differences in group delay and Doppler
frequency between the two scattered waves. The Doppler shift 
depends on the relative velocity between the source and 
observer along the (scattered) lines of sight. Hence the 
\it difference \rm in Doppler shift of waves scattered 
along two different paths through the ISM depends
only on the angles of scattering and the \it transverse \rm 
velocity of the observer and pulsar relative to the 
ISM. It is the beating of these slightly different 
Doppler shifts that makes the fringe vary in time 
and so their difference equals the fringe rate $f_t$.  
There is a quadratic relation of delay to fringe rate 
because, while the delay depends on the square of the 
angle of scattering, the Doppler shift depends linearly
on the angles.  When one of the waves is unscattered the 
result is a simple parabola $f_{\nu} \propto f_t^2$.

The observational properties
of pulsar scintillation arcs have been described by
Hill et al.\ (2003), who observed arcs over a broad
range of wavelengths and found the curvature
to be remarkably constant and to follow the 
expected scaling as wavelength squared.
Arc observations from five pulsars are shown by
CRSC, who summarize the various arc phenomenon.  
The results show many deviations from
the simple arc $f_{\nu} \propto f_t^2$.
A particuarly striking form is the 
occasional appearance of ``reverse arclets''
which are displaced from the origin and whose curvature 
is equal in magnitude, but reversed in sign, to 
that of the main (forward) arc.
Hill et al.\ (2005) reported on a set of remarkable
stable reverse arclets from PSR B0834+06.

We present results for the evolution of the arcs 
seen in PSR~B1133+16 at 408 MHz in archival 
observations spanning 16 months.  On two of the 
days observed with short time resolution
the delay-Doppler spectrum showed 
several discrete arcs with reversed curvature, 
similar to those shown by Hill et al. (2005). 
We concentrate on fitting a model to one of those days.
From an assumed brightness distribution we calculate 
the expected delay-Doppler spectrum using the screen theory of 
CRSC. Then we iteratively adjust the parameters of the 
model to fit to the observed spectrum. 
Although the fitting is not unique we are able to explain the 
reversed arcs and obtain an estimate of the scattered 
brightness distribution.  This model
distribution is then compared with that expected 
from a Kolmogorov scattering medium. 
We finish with a brief discussion of what
structures in the interstellar medium might be
responsible.

The technique we use, which is model-fitting to match 
the observed secondary spectrum, differs 
from the inversion described by Walker and Stinebring 
(2005), who perform the match to the primary dynamic 
spectrum.  We discuss the differences between the two
methods in \S\ref{sec:theory}.

%====================================================
\begin{figure*}[htb]
\includegraphics[height=15cm, angle=270.0]{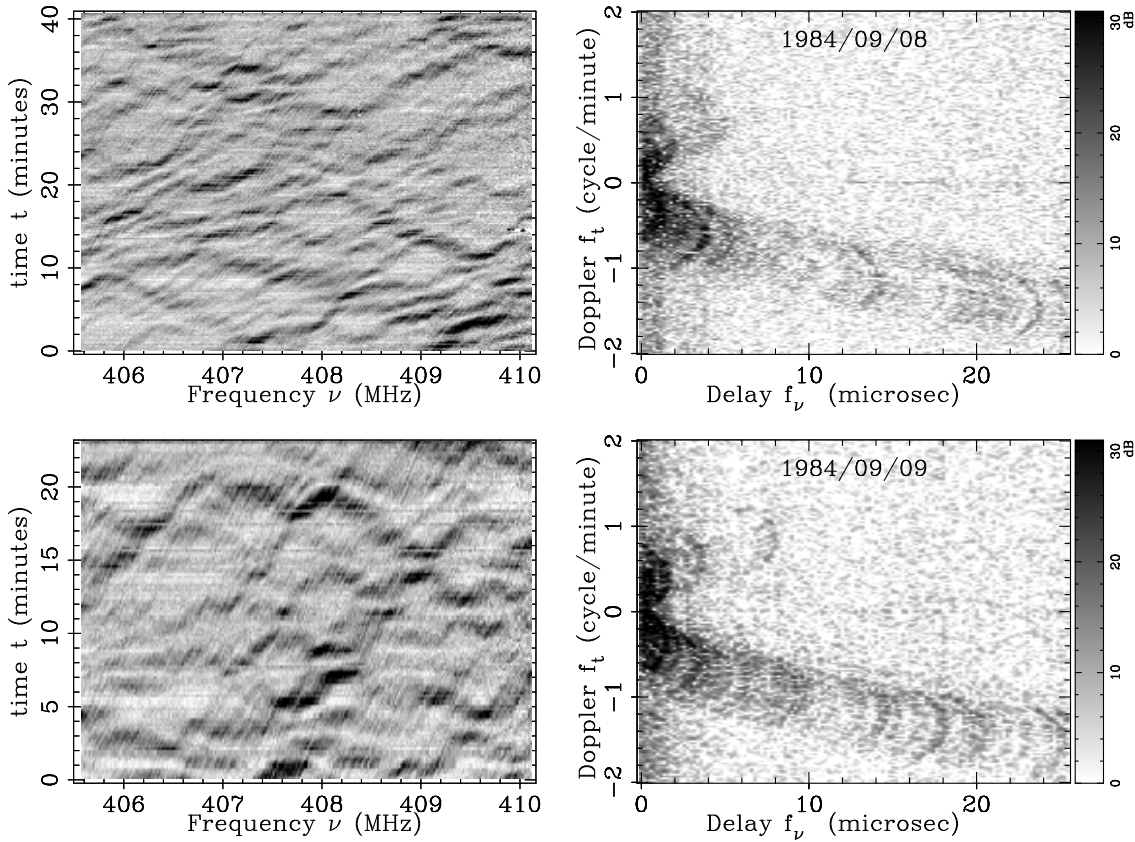}
\caption{\it Top Left: \rm  Dynamic Spectrum of PSR~B1133+16
on 8th September 1984 (day 0) with sampling time of 10 seconds.  
\it Top Right: \rm Secondary Spectrum of the same data. 
The plot is limited to $\pm2$ cpm in $f_t$, since 
no signal is visible in the range 2-3 cpm.
The grey scale is logarithmic from the mean noise 
level to black at a level 30dB higher, as shown by the
tablet to the right. The plot
shows a set of reversed arcs grouped together 
forming a forward parabolic arc predominantly visible
for negative $f_t$.
\it Bottom Left and Right: \rm  Same displays for 9th
September 1984.  Note that the different time scale 
from that above;  both show narrow fringes that drift in the same
general sense.  Both secondary spectra
are similar, but the details of the reverse arcs
have changed over one day.}
\label{fig:days01}
\end{figure*}

\section{Observations and Data Analysis}
\label{sec:obs}

The observations were made with the 76m telescope at 
Jodrell Bank between September 1984 and January 1986, 
as described by Gupta, Rickett, \& Lyne (1994).  
An auto-correlation spectrometer was centered 
at 408 MHz with 256 channels of 20 kHz 
covering 5 MHz. The spectrometer was gated 
synchronously with the pulsar period and 
spectra were averaged in windows on and off 
the pulse. The averaging time was normally 
30 seconds and on two occasions it was shortened 
to 10 seconds.  The duration ($T_{\rm obs}$) 
of most observations was from one to four hours. 
The on and off pulse spectra were corrected for 
the one-bit auto-correlation normalized and 
subtracted to create a primary dynamic spectrum 
of pulse intensity $S(\nu,t)$.

The left hand panels of Figure \ref{fig:days01} shows 
%the primary dynamic spectrum 
$S(\nu,t)$ for 8 and 9 September 1984
(which we call days 0 and 1 of the observing sequence);
they exhibit crisscross structure with a 
predominance of scintles that drift toward
higher frequency with time. With the 10 second sampling
one can see fine scale fringes drifting in the same direction.

The right panels are the corresponding secondary spectrum
-- $S_2(f_{\nu},f_{t})$ or  delay-Doppler spectrum, which 
is the squared magnitude of the 
Fourier Transform of the primary spectrum: 
\begin{eqnarray}
S(\nu,t) & \longrightarrow & S^{\dagger}(f_{\nu},f_{t})\\
S_2(f_{\nu},f_{t}) & = & |S^{\dagger}(f_{\nu},f_{t})|^2 \; .
\end{eqnarray}
Here $S^{\dagger}(f_{\nu},f_{t})$ is the two 
dimensional discrete Fourier transform of the 
sampled primary spectrum $S(\nu,t)$.  
%Its arguments are conjugate time, $f_{t}$, and conjugate 
%frequency, $f_{\nu}$, or delay and Doppler shift, respectively. 
With time resolution of 10 seconds and frequency resolution of 
20kHz, the Nyquist frequencies are $f_{t,{\rm Nyq}}=3$ 
cycles min$^{-1}$ (cpm) 
and $f_{\nu,{\rm Nyq}} = 25.6$ cycles/MHz (or 25.6 $\mu$sec).  
The associated spectral resolutions are 
$\delta f_t = 1/T_{\rm obs}$ and $\delta f_{\nu}=0.2 \mu$sec.

The $S_2$ plots show structure
which is strongest near the origin and extends out
to large delays ($f_{\nu}$) in the lower part of the panel
(ie for negative $f_t$). We call this the forward arc,
since it is curved toward the $f_{\nu}$ axis.
It consists of discrete curves, which we call reverse arcs,
since they are curved in the opposite sense.  
The apexes of the reverse arcs lie approximately 
along the parabola $f_{\nu} \propto f_t^2$.  
On close inspection one can see several 
nearly parallel forward arcs. The top and 
bottom panels of Figure \ref{fig:days01}
are separated by 27 hrs and show very similar behavior
in general, but the individual reverse
arcs have clearly changed substantially.
The main goal of our paper is to investigate
these reverse arcs and to find a model for the
interstellar angular scattering function that could 
explain them.

%--------------------

\begin{figure*}
\includegraphics[width=13cm]{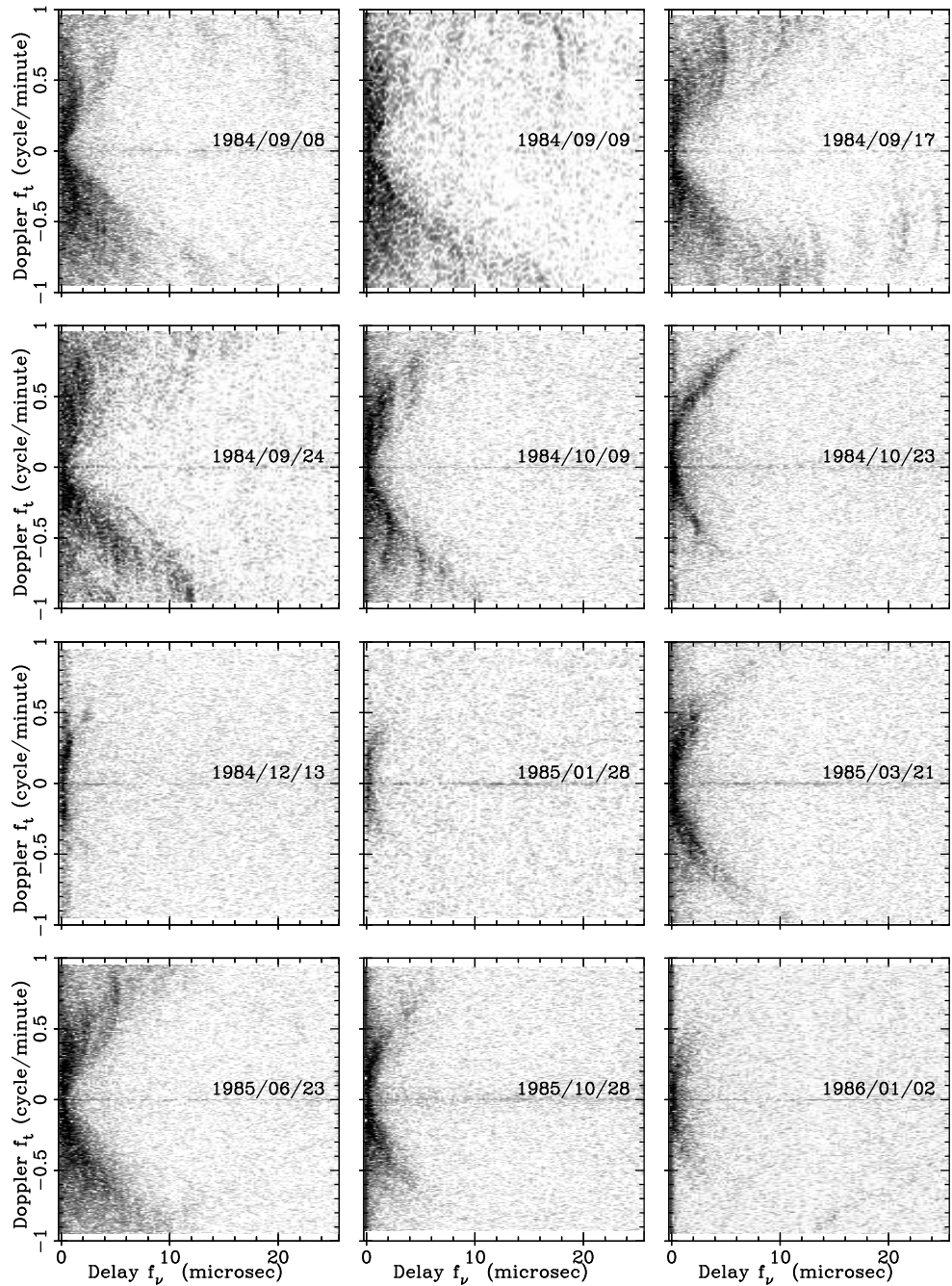}
\caption{Selected Secondary Spectra of PSR~B1133+16 from 
September 1984 to January 1986 in a logarithmic grey scale
over a 30 dB dynamic range, as shown by the tablet in Figure
\ref{fig:days01}
}
\label{fig:full12}
\end{figure*}

\subsection{Time Evolution of the Arcs}
\label{sec:evol}

After rapidly sampled observations were made
on September 8th and 9th 1984, a slower standard
sampling interval of 30 seconds was adopted
for monitoring the ISS of PSR~B1133+16.
Figure \ref{fig:full12} shows $S_2$ plots from 
twelve observations spanning 481 days. 
Days 0-31 show the same general character
of a V-shaped ``valley'' with little power
along the delay axis,
flanked by diffuse power often appearing 
as a set of discrete reverse arcs, which 
can be asymmetric in $f_t$.  
Then day 45 shows a simpler forward
arc which fades remarkably over days 96 and 142.
A strong forward arc appears
on day 194 and by day 288 multiple reverse
arcs similar to days 0-31 have returned.
Day 415 is similar to day 194, but day 481
has weak and indistinct arc structure.
These remarkable changes in the appearance of the secondary spectra
are evidence that the line of sight
crossed through particular localized structures,
as we discuss in \S~4.

With 30 second sampling
$f_{t,{\rm Nyq}} = 1$~cpm, and so features 
beyond $\pm 1$~cpm can be aliassed into
the figure.  This is evident by comparing 
the plots for days 0 and 1 in figures (\ref{fig:full12}) 
and (\ref{fig:days01}).
For example on day 1 figure (\ref{fig:full12}) shows
a linear feature at $f_t\sim 0.75\pm0.15$ cpm
and $f_{\nu} \sim 18 \mu$sec.  This 
is an alias from $f_t\sim 0.75-2.0 = -1.25$ cpm,
visible in the lower right panel of Figure \ref{fig:days01}
at  $f_{\nu} \sim 18 \mu$sec with
its apex near $f_t = -1.4$ cpm.
The figure (\ref{fig:days01}) data were a 23 minute 
observation immediately following the 30 min
observation figure (\ref{fig:full12}). 
This discrete reverse arc changed very little in 40 minutes. 
Though it clearly changed between day 0 and 1.
Days 9 and 16 show fuzzy reverse arcs 
for both positive and negative $f_t$, which
are probably confused by aliasing, but give evidence
for an evolution in the asymmetry
as well as in the reverse arc features.

In one of the observations of PSR B1133+16 shown by CRSC
the secondary spectrum has two very well defined discrete forward arcs.  
Putney and Stinebring (2006) have found multiple arcs
in six pulsars.  For B1133+16 they found the arc curvatures 
to fall near four discrete values, but that the arc 
strengths were quite variable.
They suggest that these are due to variable amplitude scattering
at four discrete distances from the Earth.
Consequently, we have estimated the curvature $a$ 
(defined by $f_{\nu} = a f_t^2$) from 
each of our observations scaled to 1 GHz.  The mean of our values 
over 1984 September to 1986 January was 
$a = 6.7\times 10^{-3}$~sec$^3$ with an rms scatter of $0.6 \times 10^{-3}$~sec$^3$ ,
which is consistent with one of their values at
$(6.3 \pm 0.1)\times 10^{-3}$ sec$^3$.  We do not detect multiple 
forward arcs, but note the signal-to-noise ratio in our data
is lower than in most of Stinebring's data.  We also note that the
curvature estimation becomes more difficult in the presence of
reverse arclets, and in such cases we trace the arc by the location of
their apexes. 

%====================================================

\section{Modeling of arcs from PSR~B1133+16}
\label{sec:model}

\subsection{Theory}
\label{sec:theory}

We use the theory of arcs in the secondary spectrum
as given by CRSC.  Section 3.3 of that paper
analyzes $S_2$ in terms of a ``brightness function'' 
$B(\theta_x,\theta_y)$ that represents the
instantaneous strength of contributions to 
the intensity at a given observing point
as a function of the angle of arrival
$\thvec=(\theta_x,\theta_y)$. The theory
provides insight into the basic interference effects
that give rise to the arc phenomenon,
but it does not include the dispersive delays
introduced by propagation through the
irregular plasma density in the ISM.
In \S\ref{sec:simulation}
we use a full electromagnetic simulation
to include such effects.

Consider a pulsar at distance $D$
whose radiation is scattered by a screen 
at a distance $sD$ from the pulsar.
The observer sees a superposition of
interference fringes from each pair of apparent 
angles of arrival $\thvec_1$ and $\thvec_2$.
The fringes contribute to a single pixel in
$S_2(f_{\nu},f_{t})$ if the angles obey:
\be
f_{\nu} &=&  \left( \frac{D(1-s)}{2\,c\,s }\right) 
\left(\theta_2^2 - \theta_1^2\right) 
\label{eq:fnu} \\  
f_t &=& 
  \left(\frac{1}{\lambda\, s}\right)(\theta_{x2} - \theta_{x1}) \viss, 
\label{eq:ft}	
\ee 
where $c$ is the speed of light and
$\lambda$ is the central wavelength observed.
$\vissvec$ is the effective transverse velocity
relative to the scattering region
of the line of sight from pulsar to observer,
which both may be moving (Cordes and Rickett, 1998);
without loss of generality we have chosen the orientation
of  $\vissvec$ to define the direction of $\theta_x$.

Equation (8) of CRSC expresses $S_2(f_{\nu},f_{t})$ 
as a four-fold integral over all pairs of angles, subject 
to the constraints of equations \ref{eq:fnu} and \ref{eq:ft}.  The constraints allow
the result to be reduced to a two-fold integral
as given by CRSC equations (9) or (B4 \& B5). 
We used the latter, since it is more convenient 
numerically, and we rewrite it here in a slightly different
form as $s_2(q,p)$ which is the delay-Doppler spectrum in terms of
normalized delay and fringe rate:
\begin{equation}
p = f _ \nu {2cs \over D(1-s) {\theta_0}^2}, 
\qquad q = f_t {s\lambda \over \viss 
\theta_0}
\label{eq:pqdef}
\end{equation}
$\theta_0$ is a normalizing angle, typically chosen to be the 
angular radius of the diffractive brightness distribution.

The result is
\be
s_2(p,q) & = & {\pi \over q}\int_{-\infty}^{\infty}\int_{-\infty}^{\infty} 
b(\xi_{+},y_1)b(\xi_{-},y_2)dy_1dy_2  \label{eq:s2int}   \nonumber \\
\xi_{\pm} & = & {p \over 2q} \pm {q \over 2} - 
{1 \over 2q}({y_2}^2 - {y_1}^2) 
%   \zeta & = & {p \over q}{1 \over 4\pi} + q\pi - {1 \over q}{1 \over 
% 4\pi}({y_2}^2 - {y_1}^2) \nonumber
\ee

where $b(x,y)$ is the brightness function in terms of
normalized angular coordinates  
$x = {\theta_x \over \theta_0},  y = {\theta_y \over \theta_0}$,
and the constraints in equations (\ref{eq:fnu} \& \ref{eq:ft})
become:
\be
p & = & ({x_2}^2 - {x_1}^2) + ({y_2}^2 - {y_1}^2) \nonumber \\
q  =  x_2 - x_1  . 
\label{eq:pq}
\ee
There is also a second mirrored relation 
obtained by mapping $p\Rightarrow -p, q\Rightarrow -q$.

Hence we can describe an arc as the locus of 
points in $p,q$ due to interference of a narrow discrete 
component in $b(x,y)$ at $x_p,y_p$ with the remaining
extended continuum in $b$, as is done in section 3.4.3 of CRSC.
A primary forward arc $p \ge q^2$ comes from
an undeviated discrete component in $b$
interfering with the extended continuum in $b$.
This is the condition under weak
scintillation described by CRSC.  The boundary of this primary arc
is $f_{\nu} = a f_t^2$ (or $p= q^2$). 
In turn $a$ is related to the physical parameters
via equations \ref{eq:ft} and \ref{eq:fnu}.

In general we substitute $x_p,y_p$ for $x_2,y_2$ in equation
(\ref{eq:pq}) to obtain a relation between
$p$ and $q$ depending on $x_1$ and $y_1$, which
vary over the continuum in $b$. The special case
where $b$ is narrow in one
dimension and extended along a line
$y_1=x_1 \tan{\phi}$, at an angle $\phi$ to the $x$-axis,
allows us to eliminate both $x_1$ and $y_1$ to obtain
the quadratic relation between $p$ and $q$:
\begin{equation}
p = x_p^2+y_p^2 - \sec^2(\phi) \;(q-x_p)^2    .
\label{eq:pqarc}
\end{equation}
This reverse arc is a parabola with an apex at
$p_a= x_p^2+y_p^2, q_a= x_p$
and negative curvature $-\sec^2(\phi)$, which
has unit magnitude if $\phi=0$ (i.e.\ scatter broadening
parallel to the velocity). The corresponding forward arc
with its apex at negative $p$ is obtained
by reversing the signs of $p$ and $q$ in
equation (\ref{eq:pqarc}). Both sets of arcs are 
shown in Figure 4 of CRSC.

Of course, in general the scatter broadening function
is not one-dimensional, as assumed above. Further,
we have not discussed the power density of $s_2$ along an arc.
This depends on the relative strength of the discrete
component supposed to exist in $b$ and on the shape 
and intensity of its continuum, as described by
the integral (\ref{eq:s2int}).

%----------------------

\subsection{Model for September 8th 1984 (day 0)}

The parallax and proper motion of PSR~B1133+16
were measured by Brisken et al. (2002);
they found its distance $D= 0.35\pm 0.02$ kpc and 
its transverse velocity $V_{\rm pm}=631 \pm 36$ km~sec$^{-1}$.
In general the scintillation velocity depends
on the transverse velocities of the pulsar, the Earth and 
the scattering medium, but since our pulsar is moving
much faster than the Earth and the medium, we have 
$\viss = V_{\rm pm} (1-s)$.  Hence the arc curvature,
as given by equation (8) of Stinebring et al. (2001),
becomes:
\be
a = \frac{D \lambda^2}{2cV_{\rm pm}^2}\frac{s}{1-s} = 0.0244 \frac{s}{1-s} \; {\rm sec}^3  \; ,
%f_{\nu,\mu s} = 6.77 f_{t,\rm cpm}^2 s/(1-s) , 
\label{eq:curv}
\ee
at our observing wavelength. 
The arc curvature $a = 6.7 \times 10^{-3}$ sec$^{3}$, 
given in section 2.1, is scaled to 1 GHz.  
When scaled back to our observed frequency of 408 MHz, 
we have $a = 0.040 \pm 0.003$ sec$^{3}$.  
This results in $s = 0.62$, giving a screen at 133 $\pm$ 7 pc from the Earth.
Putting the mean value into equations \ref{eq:fnu} and \ref{eq:ft}
gives the mappings:
\be
f_{t,\rm cpm} = 0.15 \,(\theta_{x2,\rm mas}-\theta_{x1,\rm mas}) \nonumber \\
f_{\nu, \rm \mu sec} = 
0.26 \, (\theta_{2\rm mas}^2-\theta_{1\rm mas}^2).
\label{eq:map}
\ee

The secondary spectrum from September 8th 1984 shown in 
the upper right panel of Figure \ref{fig:days01}
exhibits a series of reverse arcs extending 
to delays of 25 $\mu$sec, which are visible predominantly
at negative $f_t$.  We choose these data to model
quantitatively, because of these unusual 
\it asymmetric discrete reverse \rm arcs.
Consider, for example, the faint reverse arc 
with its apex at $f_{\nu} = 14 \,\mu$sec and
$f_t = -1.1$ cpm, which is visible
over the approximate range $-1.6 \le f_t \le -0.8$ cpm.

A discrete reverse arc
is formed by interference of a point component
in the scattered brightness function 
with an anisotropic extended component. To simplify the analysis
consider the normalized brightness $b(x,y)$ to be narrow
in $y$ and extended in $x$, i.e.\ parallel
to the ISS velocity (i.e. $\phi=0$ in equation(\ref{eq:pqarc})).  Thus we write 
$b(x,y) = [a_p \delta(x-x_p) + b_e(x)]\delta(y)$,
where $b_e$ represents the extended component
and we use delta functions to model the narrow
structures.  The interference between waves from
$x_p$ and $x$ gives $q=x_p-x$. Thus, if
$b_e(x)$ extends from $x_{-}$ to $x_{+}$, the 
reverse arc will have its apex at $p_a = x_p^2$, 
$q_a=x_p$ and be visible over
$x_{-}-x_p \le q \le x_{+}-x_p$. 

In this model $y=0$ and the apex lies on the basic arc. Hence
substituting $f_{\nu}$ and $f_t$ from above into equation
(\ref{eq:curv}) gives $s=0.63$, which is consistent
with the value 0.63 obtained above.  Using the mapping
in equation \ref{eq:map},
the observed range $-1.6 \le f_t \le -0.8$ cpm
for the reverse arc, corresponds to
interference of a discrete component at
-7.5 mas interfering with an extended component
from -2.7~mas to 2.7~mas.  In a similar way for each discrete
reverse arc on day 0 we can find a corresponding
pair of components in the scattered brightness
(a point and extended component along $\theta_x$).

Following this idea we model the brightness 
distribution (in normalized coordinates)
as a sum of Gaussian functions:
\begin{equation}
b(x,y) = \sum_{i=1}^n A_i e^{- {x-x_{oi} \overwithdelims () \sigma_{xi}}^2 - 
{y-y_{oi} \overwithdelims () \sigma_{yi}}^2}
\label{eq:modeleq}
\end{equation}
where $A_i$ is the intensity, $\sigma_{xi}$ describes the width 
of $x$, $\sigma_{yi}$ describes the width of $y$, $x_{oi}$ 
is the offset in $x$, and $y_{oi}$ is the offset in $y$. 
We made an initial model in the manner described above
for the most obvious reverse arcs in Figure \ref{fig:days01}.
As more components are added the extended component 
eventually  becomes the sum of the discrete components.
Then we tried to optimize the model by
minimizing the sum of the squared errors between 
the observed $s_2$ and the model obtained by a numerical integral of
equation (\ref{eq:s2int}).   

\subsection{Fitting Process}

We chose to work in normalized delay $p$, fringe rate $q$
and angle $x,y$ and used a normalizing angle $\theta_0 = 3.1$~mas, 
with the constants evaluated as discussed above.   
The fitting was done over two regions where the arcs 
are visible in the top right panel of Figure \ref{fig:days01}:
a lower rectangle  -2--0 cpm by 0--25 $\mu$s
and a smaller upper rectangle  0--1 cpm by 0--5 $\mu$s.

We estimated  the noise background in $S_2$ from the 
average noise level in $S_2$ over the regions not included 
in the fit, and this constant 
%$y_{noise}$ 
was added to the model.
We defined $\chi^2$ in a standard fashion: 
\be
\chi^2 = \sum^{i,j} {{[S_2(f_{\nu,i},f_{t,j}) 
- S_{2,{\rm mod},i,j}]^2}\over{\sigma_{i,j}^2}} , 
\ee
where $\sigma_{i,j} = S_{2,{\rm mod},i,j}$. At each point $S_2(f_{\nu,i},f_{t,j})$ is an
estimate of the secondary spectrum, which has a standard deviation
$\sigma_{i,j}$ equal to its mean value (following exponential statistics).
Once the model is close to an optimum value it provides an estimate 
of $\sigma_{i,j}$. Thus we used the model as an estimate 
for $\sigma_{i,j}$ rather than the observed value.

The fitting was done in a semi-automated fashion.
Starting with separate optimization of the parameters describing each
narrow component responsible for the more obvious discrete arcs, 
we minimized $\chi^2$ over a rectangle around each apex. 
Table \ref{table:model} lists the coordinates of the apexes of six of the most
obvious reverse arcs. 
With five parameters for each Gaussian the number of parameters to fit
grows quickly as components are added.  To simplify we started
with all components centered on the x-axis (ie $y_{oi}=0$).
With the position $x_{oi}$ of a component fixed we explored the 
influence of $\sigma_{xi}$ and $\sigma_{yi}$.  For the arcs at large
delays we found that $\sigma_{xi}$ had to be quite narrow
to agree with the observed arc thickness in delay, 
but that $\sigma_{yi}$ could be substantially wider.
As a result we did not optimize for the $y$-position 
or width, but set them all on the $x$ axis with $\sigma_{yi}=0.3$~mas.
This left three parameters to be optimized for each component
(amplitude, $x$-position and width).  However,
as discussed above the extent in $q$ is governed by the $x$-
position and width of the associated extended component, 
thus they too had to be adjusted.

Close inspection of the observed $S_2$ shows narrow
parallel forward arcs crossing the reverse arcs.
As can be seen from Figure 4 of CRSC, these are
extensions of the reverse arc to positive $f_t$ values
which are reflected about the origin.  Thus they show
interference of a discrete component ($x_p <0$) with components of
the scattered brightness extended over a wide 
range, extending to $x$ values more negative
than $x_p$.  This emphasizes that in fact
the entire set of arcs is caused by the mutual
interference of all possible pairs of points in $b(x,y)$.

As noted above the observations show parallel
forward arcs (for negative $f_t$), but they also
include some that appear as fainter (whiter) stripes.
Such fainter stripes come from a discrete
minimum in $b(x,y)$, whereas the brighter
arcs are from discrete maxima.  We modelled this
effect by putting dips into the brightness
function, which we achieved by
having two offset broad extended components with a
minimum in $b$ between them. These minima improved 
the fit slightly over a model without the minima.
Thus in choosing how many
and where to place new components, we were guided by the
visual appearance of the observed $S_2$
in comparison with the model. 

Table \ref{table:parameter} gives a list of the parameters for the 9-component
model for $b(x,y)$ (scaled into milli-arcseconds).
Components 1-6 are relatively narrow
($\sigma_x < 0.3$~mas) and
components 7-9 are relatively wide.
It is the interference of the narrow with the
wide components that is responsible for the 6
most obvious discrete reverse arcs.  
Their position and width in Doppler frequency
governs the range in $x$ of the brightness function.

We optimized the model to minimize $\chi^2$ 
with regard to the three variable 
parameters of the 9 components. The model
can be seen in the upper panel of
Figure \ref{fig:ss-model},
to be compared with the upper right panel of
Figure \ref{fig:days01}.
However, the value of $\chi^2$ was about five times 
the number of points in the fit, indicating that
the model is far from optimum even though
it was a local minimum. The largest discrepancy was 
traced to the region near the $f_t$ axis
(from -1 to -2 cpm), in which the observed power level
is elevated but the model is not.  The higher
level in the observations appears to be due
to effects from broadband variation
of the pulse strength (intrinsic
pulse modulations or impulsive interference).
Consequently, we estimated this extra power
by averaging $S_2$ in (from -2 to -3 cpm)
as a function of delay, and added it to the model.

This refinement caused little change in the best parameters
but reduced the minimum $\chi^2$ to about twice the number
of data points.  Visually, the model
resembles the observations in the 
asymmetry versus $f_t$, the presence of
6 reverse arcs, the location of their apexes,
and their width in delay ($f_{\nu}$). 
Given the success of the fit from a visual perspective,
we decided not to attempt a better fit by
introducing extra components, since 
we would just be refining the model for a particular
snapshot of the scattering function $b$.
Further, since our method of fitting relies on the visual
impression in choosing how many components and their
initial parameters, the result is not unique.

The upper panel of Figure \ref{fig:ss-model} shows the 
model secondary spectrum. It corresponds
to the brightness model in Table \ref{table:parameter}
with one modification.  In order to represent the
``speckle'' effect in a snapshot of the scattered brightness
we multiplied the sum of Gaussians by 
a random function of $x$ (uniform distribution), before computing
the integral in equation (\ref{eq:s2int}).  
The effect is to cause a set of fine nearly parallel
forward arcs, somewhat reminiscent of the 
observed fine structure in $S_2$.
The model brightness is plotted logarithmically
in the upper panel of Figure \ref{fig:bright-model}.
Note the asymmetry and anisotropy in its
distribution of  discrete components along the
axis parallel to $\vissvec$.  

The effect of rotating the 
brightness by an angle $\phi$ to the $x$-axis 
increases the magnitude of the curvature
in $p,q$ plane, as can be seen
from equation (\ref{eq:pqarc}).  However,
this could be compensated by a change in the
screen distance parameter $s$. Thus we cannot
determine both $s$ and $\phi$ and leave our results
displayed with $\phi=0$. 

We now consider what constraint can be placed
on the $y$-extent of the brightness function ($b(x,y)$).
When $y_p$ and $y_1$ are non-zero, we obtain the 
a version of equation (\ref{eq:pqarc}) in which
$y_p^2$ is replaced by $y_p^2-y_1^2$.
Let the extended and point components be centered on the $x$-axis
with widths $\sigma_{y}$ and $\sigma_{yp}$, respectively.
Then these ranges broaden the
thickness of the reverse arc in delay (ie $p$ is broadened 
by the larger of $\sigma_{y}^2$ and $\sigma_{yp}^2$.
The widths of the reverse arcs in the upper right 
panel of Figure \ref{fig:days01} are about $\pm0.05 \mu$sec.
With the scalings from equation (\ref{eq:map}), this
constrains the rms $y$-widths to be less than about 
0.44~mas.  We set them (somewhat arbitrarily) to be 0.3~mas.

\subsection{Discussion of the Model}

Although the parameters we used gave us arc and subarc 
structures that resemble the particular PSR~B1133+16 data, 
we do not imply that it is the only set nor is 
it the best set.  We have produced one model
for the brightness distribution that reproduces 
the visual form of the observations and is optimized
within a limited range of the parameters. 

From this we conclude that the asymmetrical form of 
the observed secondary spectrum and the presence of
narrow discrete reverse arcs can be successfully explained 
by an asymmetrical and anisotropic brightness function with
a number of narrow discrete features.  
The constraint on the widths in $\theta_y$ is important since, 
with an $\theta_x$-width of about $\pm 5$~mas, it constrains 
the effective axial ratio of the scattered 
brightness function to be greater than about 10.
However, an interesting feature
of the model is that the individual features
can be wider in $\theta_y$ than in $\theta_x$; in other words
they may be anisotropic in the orthogonal direction.

As shown by CRSC, an absence of power along the 
delay axis in the secondary spectrum
is caused by anisotropic broadening, enhanced
along the direction of the scintillation velocity.  
All of 12 of the secondary spectra plotted
in Figure \ref{fig:full12} show no 
power along the delay axis. Hence
the same basic anisotropy in the scattering
persisted over about 500 days, but the detailed
form of the scattered brightness changed 
substantially over days and weeks.

%==========================================================

\subsection{Screen Simulation}
\label{sec:simulation}

CRSC used the simulation code of  Coles et al.\ (1995)
to simulate secondary spectra of waves scattered
by a dispersive phase changing screen.
In normal usage the method synthesizes
one realization of a stationary random phase screen,
which modulates a plane wave. The emerging complex field
is then propagated a distance $z$ to the observer, 
by transforming into wavenumber and multiplying by the appropriate
(Fresnel) filter and back transforming to the spatial domain.
The intensity is saved along a straight line track through
the two-dimensional complex wave field.
The result simulates a time series as an observer moves
relative to a scattered wave field.
A dynamic spectrum is a set of such time series, created as
the radio frequency of the incident wave is incremented.
The secondary spectrum is computed in the same
way as for the data. 

CRSC simulated the screen as a single two-dimensional 
realization of a stationary random process following a 
given wavenumber spectrum.  They showed results for a 
Kolmogorov screen. When anisotropy was included, some 
results had multiple discrete
forward and reverse arcs.  However, the results 
differed from our observations of PSR~B1133+16
in that the discrete arcs were considerably
finer and distributed symmetrically in fringe rate
($f_t$).  The addition of a linear gradient in the 
phase across the screen (i.e.\ a simple refraction)
caused the secondary spectrum to be
asymmetric in fringe rate. They showed that the phase
gradient needed to be large enough to shift the scattered image by
at least its own radius.  This is a much larger
gradient than is expected randomly 
for a medium with a Kolmogorov spectrum.

\begin{figure}[hbt]
\begin{tabular}{c}
\includegraphics[height=5cm]{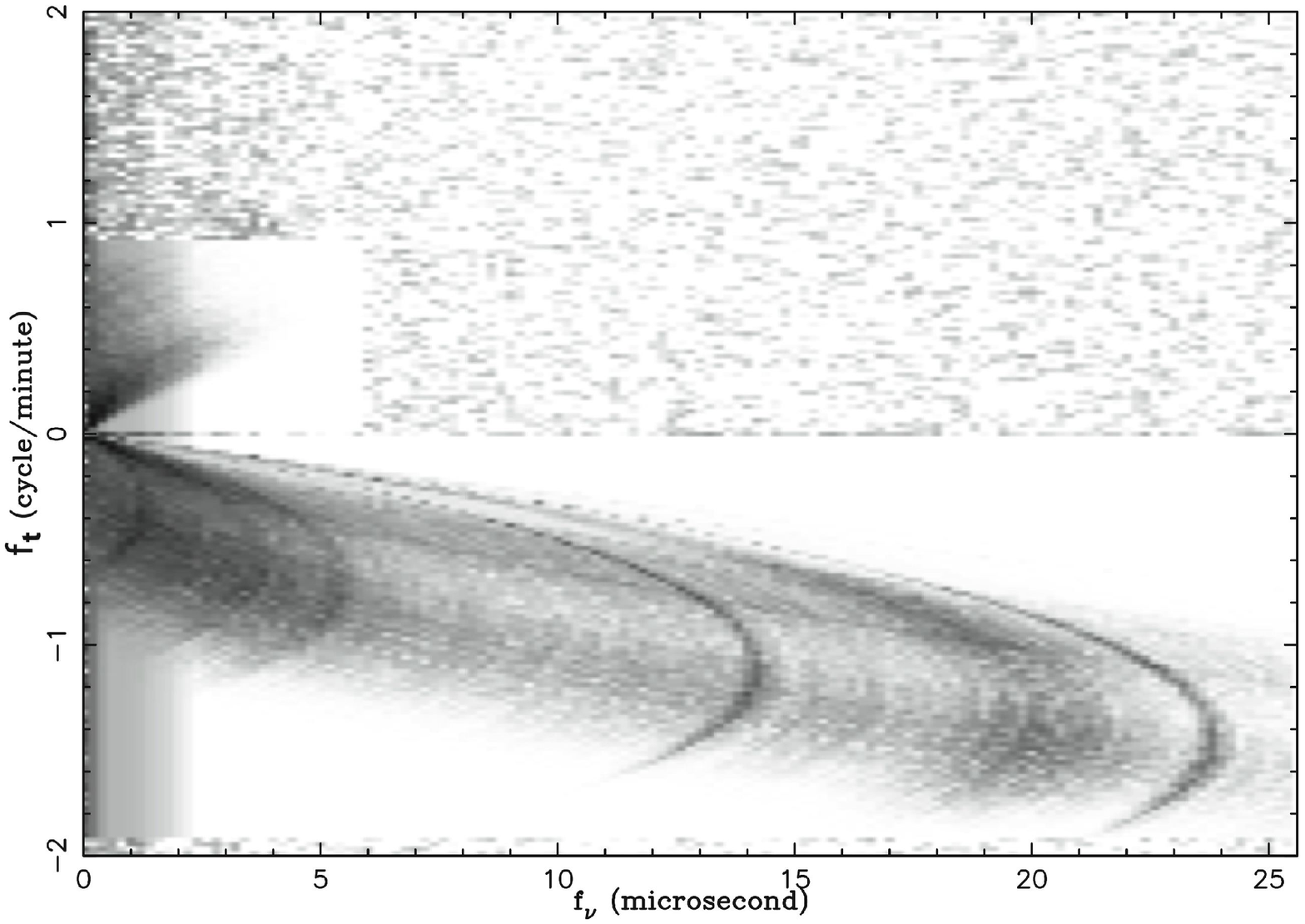} \\
\includegraphics[height=5cm]{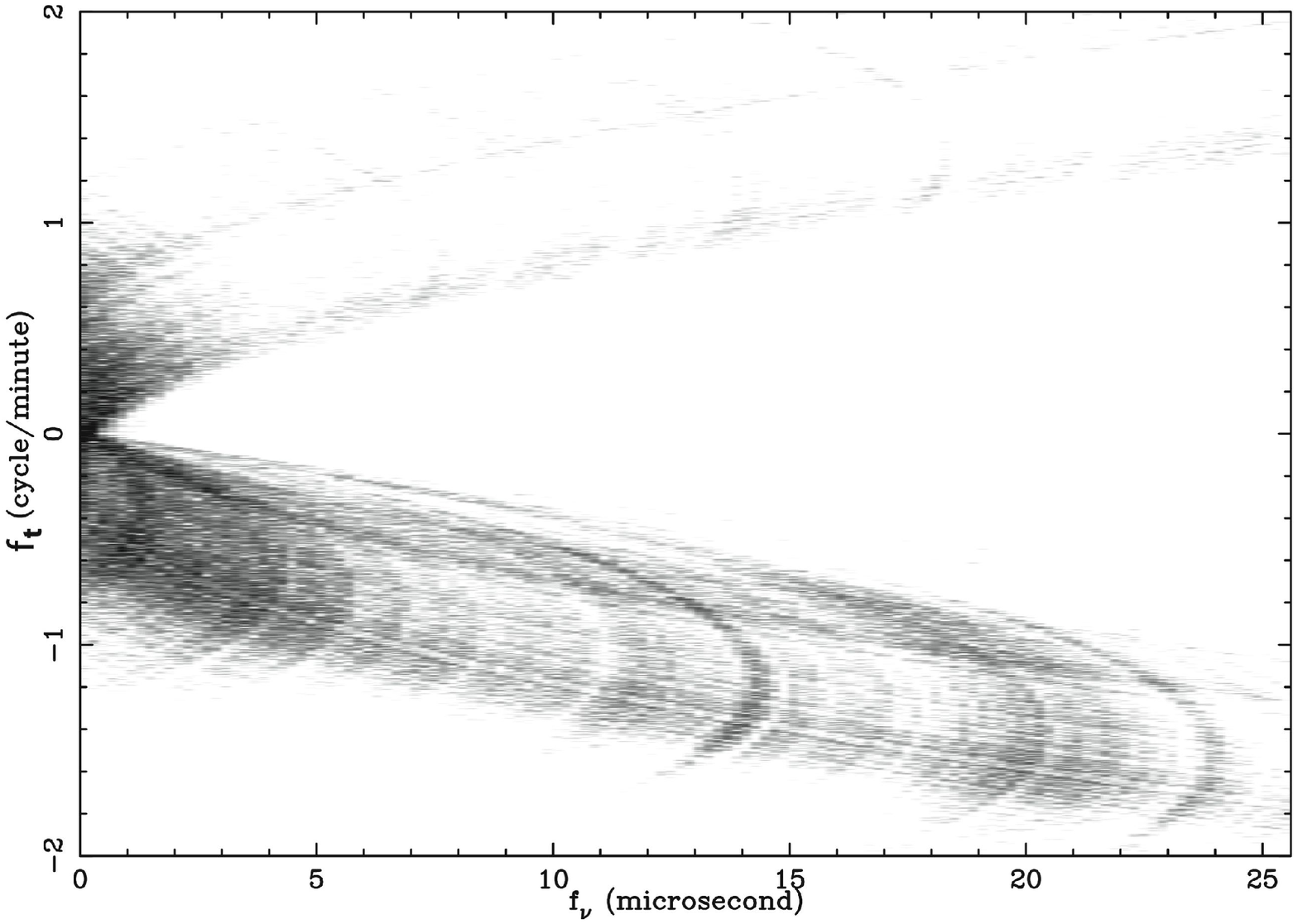}
\end{tabular}
\caption{ \it Top\rm  : Secondary spectrum computed from
the brightness distribution of the upper 
panel of Figure \ref{fig:bright-model},
by a numerical integration of equation
(\ref{eq:s2int}). Grey scale
is logarithmic with dynamic range of 36 dB. 
The fit was done over the two rectangular regions
shown; outside those regions the noise from the 
data is plotted.
\it Bottom\rm  : Secondary spectrum from the 
screen simulation for the same day (see text)
with dynamic range of 30 dB.  
}
\label{fig:ss-model}
\end{figure}

We modified the screen generation code to 
make the scattered brightness like that in our model 
(i.e.\ the upper panel of Figure \ref{fig:bright-model}).
This modification required an iteratative process,
as follows.  The field emerging from the phase screen
at position $\svec$ has unit amplitude
and is phase modulated $F(\svec) = \exp[j\phi(\svec)]$.
Its two-dimensionsal Fourier transform $F^{\dagger}(\kappavec)$
is the  spectrum of the field versus wavenumber $\kappavec$,
which maps simply to the scattered angle as
$\kappavec = k \thetavec$ with $k=2\pi/\lambda$.
Our goal is to find
a screen phase $\phi(\svec)$, for which the squared 
magnitude $|F^{\dagger}|^2$ follows our model for the 
scattered brightness $B(\thetavec)$.
We started by specifying the magnitude 
$F^{\dagger}=\sqrt{B(\thetavec/k)}$ and assigned it random phase.
Then back transformed to give $F(\svec)$, which had
both amplitude and phase modulation. A revised 
$F(\svec)$ was obtained by normalizing each point 
to unit magnitude. That revised $F$ was re-transformed
and the resulting $F^{\dagger}$ was normalized
to make its magnitude $\sqrt{B(\thetavec/k)}$
and keep its phase. When this cycle was repeated
the amplitude modulation in $F(\svec)$ is decreased.
We iterated until the residual amplitude modulation
no longer decreased ($\sim 100$ cycles).  The phase
of this final $F(\svec)$ was then extracted as the screen
phase for the normal simulation.  This method is based on the concepts
developed by  Gerchberg and Saxton (1972). 
As in the normal simulation this screen phase
was scaled proportional to radio wavelength,
to represent the plasma dispersion.

The lower panel of Figure \ref{fig:ss-model}
gives the resulting secondary spectrum.  It has
the same basic features as the simple model
in the upper panel, but shows more pronounced 
substructure in the form of fine arc modulations.
The importance of this result is that the basic
arc forms are unchanged by the inclusion of 
dispersion in the screen phase, whereas dispersion is not included
in the physics of equation (\ref{eq:s2int}).  A further result
is that we have obtained a realization of the phase
screen itself.  This is then an estimate of
the phase imposed by the ISM during the observations
on our day 0, and we can ask what it tells about
the density structure in the ISM.

\subsection{Comparison with Reverse Arcs from PSR B0834+08}

Hill, et al. (2005) reported how four discrete 
reverse arcs observed from PSR B0834+16 moved along 
the main forward parabola over 25 days.  
The Doppler frequency of the apex of each arc followed a 
linear track against time; furthermore the slopes of all 
four tracks were equal to the slope predicted by assuming that each
reverse arc was scattered from a fixed structure 
in space scanned by the known pulsar proper motion.
Their results suggest that the reverse arcs appearing on successive days 
are not unrelated, but are in fact due to the same physical structures.  

We have attempted to test this idea with the reverse arcs
observed on day 0 and 1 of our data (PSR B1133+16), which are
27 hours apart.  For the 6 most
obvious reverse arcs on day 0 we calculated 
the shift expected due to the pulsar proper motion in 27 hours
and compared the new locations to those observed on day 1.
The expected angular shift is $\delta \theta_x \sim \delta t V_{\rm pm}/D$,
which we substitute into equation \ref{eq:map}, and tabulate 
the $f_t$ and $f_{\nu}$  positions before and after this 
proper motion shift (see Table \ref{table:proper motion}).   Comparing the 
shifted coordinates with those observed on day 1, 
we found some agreement. For example, the easily visible reverse 
arc no. 4 from day 0 should should move to $f_t$ = -1.25 cpm 
and $f_{\nu}$ = 18.75 $\mu$s which is close to a prominent
arc on day 1 (lower right panel of Figure \ref{fig:days01}).
The comparison is complicated by the fact that day 1 shows a higher density
of reverse arcs making it hard to recognize particular structures.
We conclude that these data are not of sufficiently high
signal to noise ratio to confirm or reject the hypothesis
that the arc positions are simply shifted by the pulsar proper
motion.

\subsection{Relationship to W-S Inversion Algorithm}

We now compare our model fitting with the method of
Walker and Stinebring (2005) (WS) who 
estimate the scattered brightness function using an
fitting algorithm to the dynamic spectrum.  
This has great appeal since it
avoids the messy task of deciding what components 
of the secondary spectrum to include in a model fit.  
Their method represents the full electric field as
a sum of ``scattered waves'', which are successively
identified and subtracted in order to match
the observed primary dynamic spectrum to the model.
It is somewhat analogous to the CLEAN algorithm in image reconstruction.

WS consider the scattered electric field $U(\nu,t)$ at frequency $\nu$ 
and time $t$ as a sum of basis functions 
$\exp[i2\pi(\nu f_t + f_{\nu}t)]$,
which they call scattered waves, though the formulation is
not derived from scattering theory.
We note that $\nu$ and $t$ are relative to the center of
the observing band and integration time and that WS use
the symbols $\tau$ for $f_{\nu}$ and $\omega$ for $f_t$.

The dynamic spectrum $S(\nu,t)$ is then modelled by $|U(\nu,t)|^2$
and WS designed a procedure to estimate the phasor
($\tilde{U} (f_{\nu},f_t)$) for each scattered wave to best match the 
model with the observed dynamic spectrum.  The results
are given as both amplitude and phase 
for each differential delay $f_{\nu}$ and Doppler $f_t$.  
They applied the method and displayed the amplitudes 
for a particular 327 MHz observation in the
sequence from PSR B0834+06 (Hill et al., 2005), in
which there were several well-defined reverse arclets.
They discuss possible applications of the method but
do not relate the results to the scatttered brightness 
$B(\theta_x,\theta_y)$.  

However, Asplund et al. (2004) used these same results 
and mapped them to the scatttered brightness by assuming that
each component $\tilde{U} (f_{\nu},f_t)$ was the result of interference
between an undeviated (reference) wave and a wave scattered at angle
$(\theta_x,\theta_y)$.  This assumption is valid
under conditions of weak scattering, as shown by CRSC,
but it does not seem justified for the 327 MHz observations
of PSR B0834+06.  As WS note, the effective reference wave
in their algorithm starts out as undeviated, but is modified
as each scattered wave component is identified.
Nevertheless the summing of the ``scattered waves''
does include the mutual interference between the components
of the basis functions as well as their interference with the reference component. 

One can show that the secondary spectrum is the square
of the auto-correlation of $\tilde{U} (f_{\nu},f_t)$,
as follows:
\be
S_2(f_{\nu},f_{t}) = | \stackrel{F_{\nu}}{\Sigma} \stackrel{F_t}{\Sigma} \;
 \tilde{U}(F_{\nu},F_t) \, \tilde{U*} (F_{\nu}+f_{\nu},F_t+f_t) |^2
\ee
Nevertheless, it seems that more work is needed on the theoretical
basis of the algorithm to consider the circumstances under which
the weak scintillation mapping into the angle domain is properly justified.  

While the method we have used is cumbersome to implement,
it is more straightforward in that our computed model 
directly sums the waves in the domain of $(\theta_x,\theta_y)$
including their mutual interference.

%=========================================================

\section{The Density structure in the Interstellar Plasma}

Most interpretations of radio scintillation
have been formulated as constraints on the wavenumber
spectrum of the plasma density.  The Kolmogorov
spectrum has become a default model, even though
several observations point to more fluctuation power
on  wavelengths longer than the size of the 
typical radio scattering disc  (a few AU).

\begin{figure}[hbt]
\begin{tabular}{c}
\includegraphics[height=8cm, angle=270]{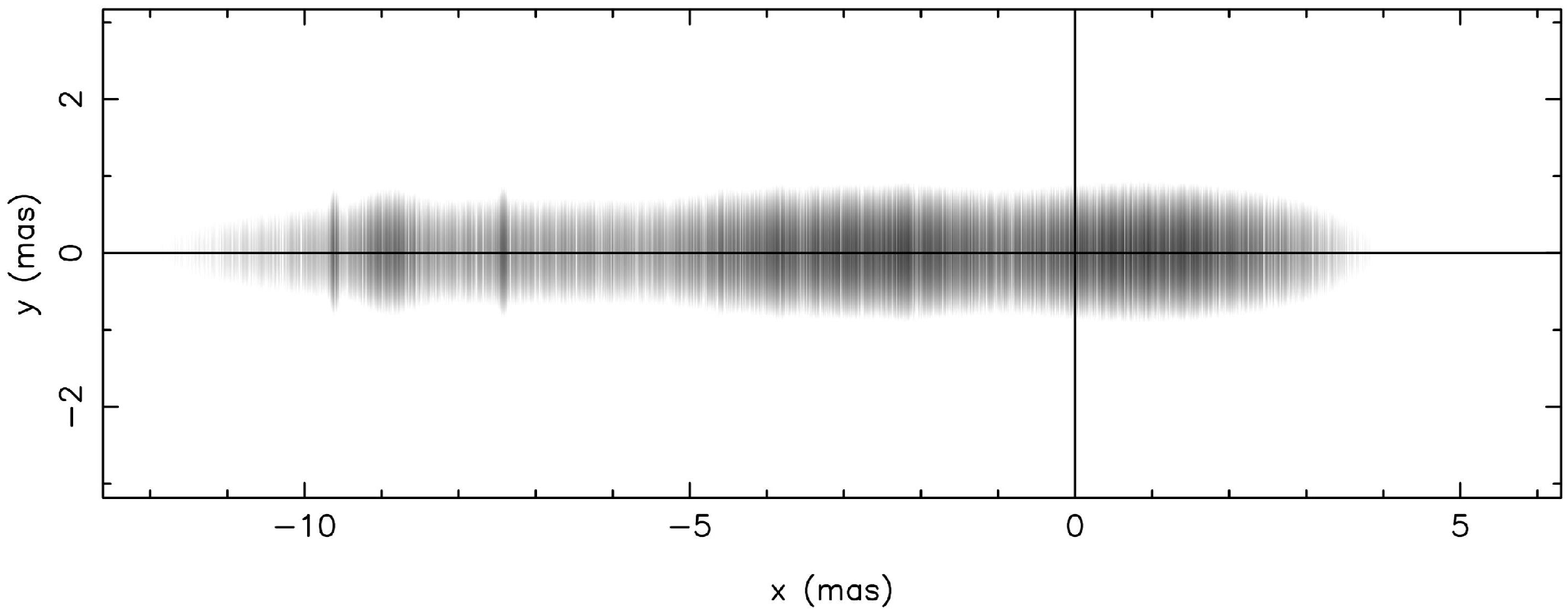} \\
\includegraphics[height=8cm, angle=270]{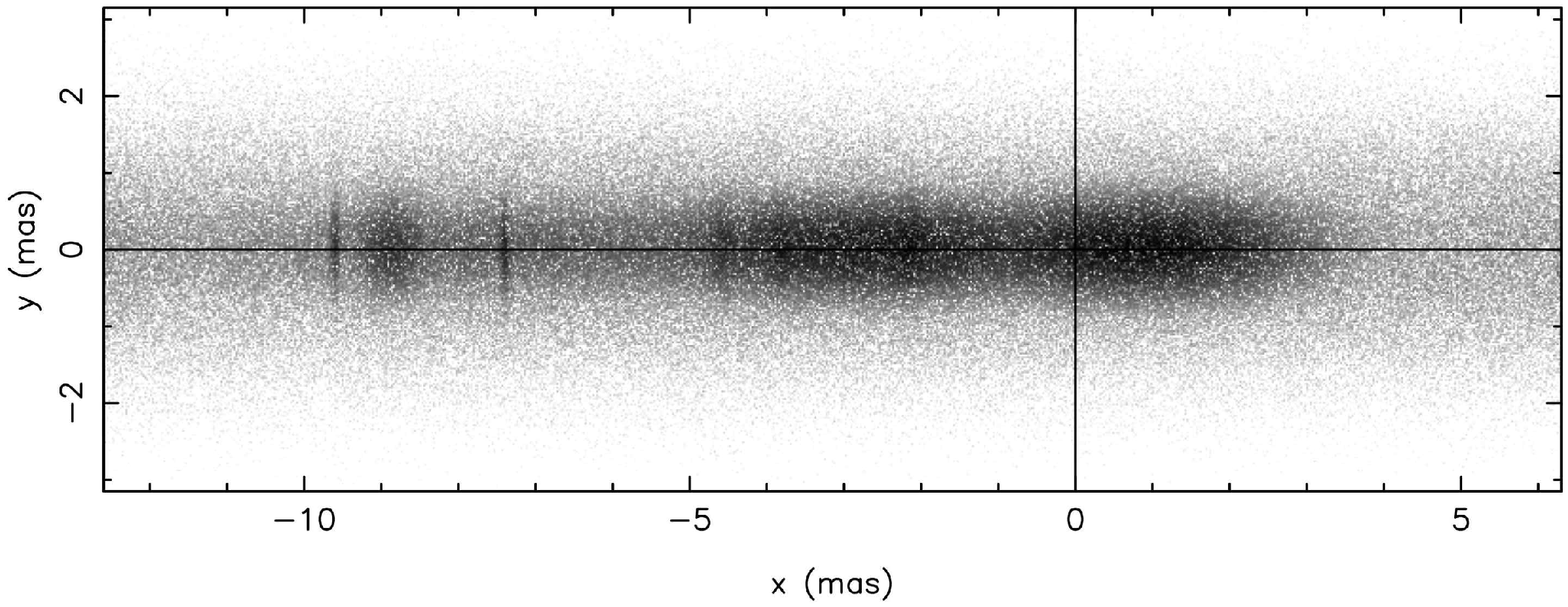}
\end{tabular}
\caption{\it Top\rm :  Logarithmic plot
of the model brightness distribution
obtained by fitting to observation on day 0
in Figure \ref{fig:days01} (dynamic range 50 dB).
\it Bottom\rm : brightness distribution obtained from the 
full screen simulation (see text).}
\label{fig:bright-model}
\end{figure}

The general concept proposed to explain the discrete
reversed arcs is that  substructure in the scattered brightness 
$b$, analogous to speckle in a scattered image,
causes substructure in the secondary spectrum,
that takes on a form of points lying on
intersecting reverse and forward
parabolic arcs, traced by equation (\ref{eq:pqarc})
and its reflection in the origin,
with intensity at each point governed by the product
of the brightness of the two interfering components.
The plots in Figure \ref{fig:bright-model} are estimates of
a single realization of the scattered brightness with a particular
sub-structure characteristic of the observing conditions.
We now consider what these images tell us about the 
density structure in the interstellar plasma.

The first point is that the scattered image is highly elongated
with an axial ratio of 10:1 or more. Further, the axis 
of elongation is roughly parallel to the 
proper motion direction of the pulsar (within, say, $\pm 20\degr$).
This means that the spatial density structures are elongated
in a direction roughly perpendicular to this direction.

The second point is that the images are not centered on the origin.
We see two possible explanations: either refraction
by a large scale transverse gradient in the column density
of electrons; or highly localized scattering caused by a 
region that is centered to one side of the line 
towards the pulsar.   

Refraction seems likely
since similar refraction has been invoked to explain the
persistent drifting in frequency of scintillation features in 
the dynamic spectra of some pulsars (e.g. Shishov, 1974; 
Hewish, Wolszczan and Graham, 1985; Gupta, Rickett and Lyne, 1994).
The basic idea is that a refracting region deflects the scattered
(and unscattered) waves which causes a lateral shift
in the scintillation pattern.  Since in a plasma the angle 
of refraction is strongly dependent on the observing frequency,
the frequency for a given maximum of intensity will drift in time.
Such a drift will be seen as an asymmetry versus Doppler frequency
in the power distribution of the secondary spectrum, this effect 
was confirmed by simulations reported by CRSC.

The alternative explanation is that there are highly localized
regions of very irregular (turbulent?) plasma, that can deflect
(ie scatter or refract) the radio waves into our line of sight.  
This implies that thay must be small enough to subtend an angle
at the observer smaller than their maximum angle of deflection.
The angles of arrival of these
waves will be determined by how far the enhanced turbulence is
from the direct line of sight and they will not depend 
systematically on frequency.

A third point about the scattered images is that they
are ``blotchy'', but the blotches have a lower filling factor
than would be expected for the usual speckle
in an angular spectrum.  (see the simulations shown by CRSC, 
in which there is a multitude of reverse arcs
due to speckle with a high filling factor from a Kolmogorov medium). 
We conclude that the blotchy nature of our scattered image is not compatible with a 
uniform Kolmogorov random medium.  Our results do not, on their
own, allow us to determine the nature of the medium.  However, the 
fascinating measurements by Hill et al. (2005) for pulsar
PSR B0834+06 provide an important constraint.  They found that
the frequency-scaling of the positions of the reverse arcs 
was not compatible with frequency dependent refraction and
implies the alternative of highly localized regions of 
enhanced scattering as the cause of each reverse arc.
Under this scenario the regions have a linear
size given by their apparent angular size times 
the scattering distance --  $D(1-s) \sim 140$~pc.
With angular extents of 0.1-1 mas their maximum implied linear sizes
are 0.014-0.14 AU.

Considering that the discrete reverse arcs were only prominent
in some of the observations (ie days 0-31 and 288), the
regions of localized enhanced scattering are irregularly distributed
in the ionized ISM.  In 31 days the pulsar moves about 10 AU
transverse to the line of sight.  Thus the persistence of 
the reverse arcs through 31 days suggests that the 
localized regions of enhanced scattering are clumped over
about 10 AU.

If localized scattering regions are indeed responsible for the 
discrete reverse arcs, the next question is what 
are these regions physically.  However, this is beyond the 
reach of our present investigation, and we refer interested readers
to Rickett, Lyne and Gupta (1997) 
and the proceedings of the workshop on
``Small Ionized and Neutral Structure in the ISM'',
held in Socorro May 2006, particularly the papers by Stinebring and Rickett.

\section{Conclusion}

The ISS of pulsar B1133+16 showed pronounced changes in 
the character of its secondary spectra in 12 days of observation 
spread over 15 months.  While these data which were centered at 408 MHz
were recorded over 20 years ago, they often show the parabolic arc
phenomenon that was discovered quire recently by Stinebring et al. (2001).  However,
the basic parabolic arc varied substantially in its extent and shape.
Discrete reverse arcs were visible for about two months early in the sequence
and then again after about nine months.  There was a two month interval when
the arc was barely detectable, corresponding to when the 
primary dynamic spectra had a very wide decorrelation bandwidth.  

We concentrated our analysis on two days with clearly visible
discrete reverse arcs, which were fortuitously sampled rapidly enough
to resolve them in Doppler frequency.   Using the theory of CRSC,
which ignores dispersion,
we modelled these features and hence were able to estimate
the scattered brightness distribution.  We tested the applicability 
of the theory, by simulating and electromagnetic wave propagation 
code that includes dispersion, and found that
the reverse arcs are not altered substantially by including dispersion.

Our estimate of the scattered brightness function is highly eleongated
with an axial ratio of more than 10:1.  It is asymmetrical about the origin
and has a small number of discrete maxima.    These properties are not consistent 
with expectations from a scattering region that behaves as
a statistically homogeneous random medium with a Kolmogorov spectrum.
When combined witth the conclusions from Hill et al. (2005), our results suggest
that there are highly localized discrete scattering regions
which subtend 0.1-1 mas, but can scatter (or refract) the radiation by
angles that are five or more times larger.  The asymmetry in 
the scattered image may well be caused by the chance configuration of these regions.

\acknowledgements  We thank Andrew Lyne for completing the series of
observations at Jodrell Bank in 1984-85 and we thank Yashwant Gupta for work on 
the original data reduction.  We thank the National Science Foundation for support under grants AST 9988398 and AST 0507713.

\begin{table}
\begin{tabular}{ccccc}
\multicolumn{5}{c}{\rule[-3mm]{0mm}{8mm}Subarc Coordinates (Observation vs. Model)} \\ \hline \hline
\rule[0cm]{0cm}{0.4cm} & Observation && Model \\
Subarc $\#$ & {$f_t$ (cpm)} & {$f_\nu$ ($\mu$s)} & {$f_t$ (cpm)} & {$f_\nu$ ($\mu$s)} \\ \hline
\rule[0cm]{0cm}{0.4cm}
1 & -0.31 & 1.2 & -0.32 & 1.22 \\
2 & -0.58 & 3.8 & -0.56 & 3.82 \\
3 & -0.69 & 5.4 & -0.67 & 5.43 \\
4 & -1.09 & 14.0 & -1.10 & 14.22 \\
5 & -1.31 & 20.0 & -1.30 & 20.36 \\
6 & -1.44 & 23.6 & -1.41 & 23.86
\end{tabular}
\caption{The coordinates of the observed individual subarcs (day 0, Figure 1: $Top$ $Right$) and the modeled subarcs (Figure 3: $Top$).} \label{table:model}
\end{table}

\begin{table}[hbt]
\begin{tabular}{cccccc}
\multicolumn{6}{c}{\rule[-3mm]{0mm}{8mm}Parameters for Model Brightness 
Distribution} \\ \hline \hline
\rule[0cm]{0cm}{0.4cm}Component \\
(i) & {($A_i$)} & {($x_{oi}$)} & {($\sigma_{xi}$)} & {($y_{oi}$)} & 
{($\sigma_{yi}$)} \\ \hline
\rule[0cm]{0cm}{0.4cm}
1 & 0.418 & -2.18 & 0.138 & 0 & 0.32  \\  
2 & 0.273 & -3.85 & 0.104 & 0 & 0.32  \\ 
3 & 0.127 & -4.59 & 0.126 & 0 & 0.32  \\  
4 & 0.327 & -7.42 & 0.041 & 0 & 0.32  \\ 
5 & 0.2 & -8.88 & 0.261 & 0 & 0.32  \\  
6 & 0.236 & -9.62 & 0.038 & 0 & 0.32  \\  
7 & 1.0 & 0.82 & 1.04 & 0 & 0.32  \\  
8 & 0.545 & -2.68 & 1.418 & 0 & 0.32  \\  
9 & 0.027 & -7.25 & 2.111 & 0 & 0.32 
%1 & $2.3 \times 10^7$ & -0.692 & 0.05 & 0.0 & 0.1 \\
%2 & $1.5 \times 10^7$ & -1.222 & 0.033 & 0.0 & 0.1 \\
%3 & $7.0 \times 10^6$ & -1.456 & 0.04 & 0.0 & 0.1 \\
%4 & $1.8 \times 10^7$ & -2.357 & 0.013 & 0.0 & 0.1 \\
%5 & $1.1 \times 10^7$ & -2.820 & 0.083 & 0.0 & 0.1 \\
%6 & $1.3 \times 10^7$ & -3.053 & 0.012 & 0.0 & 0.1 \\
%7 & $5.5 \times 10^7$ & 0.260 & 0.33 & 0.0 & 0.1 \\
%8 & $3.0 \times 10^7$ & -0.850 & 0.45 & 0.0 & 0.1 \\
%9 & $1.5 \times 10^6$ & -2.300 & 0.67 & 0.0 & 0.1
\end{tabular}
\caption{Components of the brightness distribution defined by equation 
\ref{eq:modeleq} where the angular units are mas.
These values were used in Figure \ref{fig:ss-model} to model
the observations of day 0} 
\label{table:parameter}
\end{table}

\begin{table}
\begin{tabular}{ccccc}
\multicolumn{5}{c}{\rule[-3mm]{0mm}{8mm}Subarc Coordinates (Day 0 vs. Proper Motion)} \\ \hline \hline
\rule[0cm]{0cm}{0.4cm} & Before && After \\
Subarc $\#$ & {$f_t$ (cpm)} & {$f_\nu$ ($\mu$s)} & {$f_t$ (cpm)} & {$f_\nu$ ($\mu$s)} \\ \hline
\rule[0cm]{0cm}{0.4cm}
1 & -0.32 & 1.22 & -0.48 & 2.78 \\
2 & -0.56 & 3.82 & -0.73 & 6.32 \\
3 & -0.67 & 5.43 & -0.83 & 8.35 \\
4 & -1.10 & 14.22 & -1.25 & 18.75 \\
5 & -1.30 & 20.36 & -1.46 & 25.72 \\
6 & -1.41 & 23.86 & -1.57 & 29.63
\end{tabular}
\caption{The coordinates of the individual subarcs before (day 0, Figure 1: $Top$ $Right$) and after the shift due to the pulsar proper motion.} \label{table:proper motion}
\end{table}

\clearpage

\end{document}